\documentclass[12pt]{article}
\usepackage{amssymb,amsmath}
\usepackage{graphicx,epsfig}
\usepackage{caption}
\usepackage{subcaption}
\usepackage{mathrsfs}
\usepackage{enumerate}
\usepackage{epstopdf}
\usepackage{xcolor}
\usepackage{float}
\usepackage{rotating}
\usepackage{tikz}
\usepackage{algorithmic}
\usepackage[section]{placeins}
\usepackage{textcomp}
\usepackage[normalem]{ulem}
\usepackage{derivative}
\usepackage{amsmath}

\usepackage{dsfont}
\usepackage{amsthm}
\usepackage{comment}
\usepackage{tabularx}
\usepackage{booktabs} 
\usepackage{hyperref}
\usepackage{upgreek}
\usepackage[margin=3cm]{geometry}
\usepackage[english]{babel}
\usepackage[authoryear,round]{natbib} 
\usepackage[linesnumbered,lined,boxed,commentsnumbered]{algorithm2e}
\RestyleAlgo{ruled}
\SetKwComment{Comment}{/* }{ */}
\SetKwComment{Comment}{/* }{ */}
\definecolor{darknavyblue}{rgb}{0.0, 0.1, 0.3}
\hypersetup{
    colorlinks=true,
    linkcolor=cyan,
    urlcolor=cyan,
    citecolor =cyan,
    filecolor = cyan
    }
          
\graphicspath{{Rips sequences/}}
 \numberwithin{equation}{section}

\begin{document}
\title{Censored Regression Models with Endogeneity: A Survey}
\author{
	Swati Shukla$^1$, Shalabh$^2$, Subhra Sankar Dhar $^{3}$\\
	Department of Mathematics and Statistics, Indian Institute of Technology Kanpur \\
	 Kanpur 208016, India\\
	 Emails: $^1$shukla@iitk.ac.in , $^2$shalab@iitk.ac.in,  $^3$ subhra@iitk.ac.in}

\maketitle
\begin{abstract}
  This survey article provides a comprehensive overview of methodologies developed to handle endogeneity in censored regression models. We emphasize their theoretical basis, and estimating methodologies. The paper summarizes the benefits and drawbacks of each technique and addresses expansions to nonlinear, panel, and high-dimensional settings, offering researchers practical assistance and identifying options for future methodological improvement in censored models with endogenous regressors.

\end{abstract}
\section{Introduction} 
Censored regression models constitute an important class of regression models in which the dependent variable is observed only within a certain threshold. This threshold can be either fixed or random, leading to fixed or random censoring, respectively. Fixed-censoring models have been widely studied in both theoretical and applied econometrics. In these models, parametric estimation techniques typically assume a known distribution for the error term. However, unlike classical linear regression, misspecification of the error distribution in censored regression models can result in inconsistent estimates of the structural parameters (see, e.g., \cite{Amemiya1979}, \cite{AMEMIYA19843}, \cite{Heckman1976}, \cite{Heckman1979}, \cite{GOLDBERGER1981}, \cite{arb1982}).

To address this issue, semiparametric approaches have been developed, which avoid fully specifying the error distribution while imposing sufficient restrictions to ensure identification of the structural parameters. A common restriction is the independence of the error term from the regressors, which facilitates the construction of consistent estimators (see, e.g., \cite{POWELL1984}, \cite{Powell1986}, \cite{SCLS}, \cite{Winsorized}). Despite their theoretical appeal, these estimators have practical limitations. For instance, the CLAD estimator of \cite{Powell1986} involves minimizing a non-convex objective function, and commonly used iterative linear programming methods can converge only to local minima rather than the global minimum (see \cite{Buchinsky1994}). To overcome these challenges, several alternative approaches have been proposed. Notably, \cite{Buchinsky1998} and \cite{KHAN200173} suggested subsample selection using nonparametric estimates of selection probabilities, though these methods are limited by the curse of dimensionality. \cite{Chernozhukov} introduced a simpler three-step procedure that is easier to implement, but it relies on parametric assumptions for the selection mechanism, which may result in inconsistent estimates if the selection model is misspecified.

Endogeneity presents an additional challenge in censored regression models, often arising due to omitted variables or measurement errors. Endogeneity leads to biased and inconsistent parameter estimates if not properly addressed. Various methods have been proposed to handle endogeneity in censored regression models. Moreover, many empirical applications involve both censoring and endogeneity. For instance, \cite{Chay} studied changes in black–white earnings inequality using longitudinal Social Security earnings records. In their analysis, education is treated as endogenous in the earnings equation, while a substantial portion of earnings records are censored at the Social Security maximum taxable limit, with censoring rates ranging from 32$\%$ to 54$\%$. Similarly, in the estimation of Engel curves, \cite{BLUNDELL200765} and \cite{CHERNOZHUKOV2015201} highlighted the endogeneity of total household expenditure due to the joint determination of total spending and its components. Censoring arises because a significant fraction of households report zero expenditure on alcohol, affecting the estimation of its budget share.

Over the past decades, several methodologies have been developed to address the joint problem of censoring and endogeneity. Classical approaches, such as the two-step control function method, provide consistent estimates under certain conditions. More recently, robust estimation techniques, including M-estimation, L-estimation, and quantile-based approaches, have been proposed to handle heavy-tailed errors, outliers, and model misspecification. Despite these advances, there is a lack of comprehensive surveys summarizing the theoretical developments, practical implementation, and comparative performance of these methods.

This article aims to provide a systematic overview of censored regression models under endogeneity. We discuss the evolution of estimation strategies, highlight their theoretical properties and limitations, and review applications in empirical research. Finally, we outline open challenges and potential directions for future research.

\section{Mathematical Framework}
Consider a sequence of paired independent and identically distributed (i.i.d.) data points 
\[
(y_{i}, \boldsymbol{x}^{*}_{i}), \quad i=1,\ldots,n,
\]
where $y_{i}$ denotes the $i^{\text{th}}$ observation on the response variable. The observed variable $y_{i}$ is related to an unobserved latent variable $y_{i}^{*}$ through censoring at a fixed non-random threshold $c$, such that
\[
y_{i} = 
\begin{cases}
y^{*}_{i}, & \text{if } y^{*}_{i} > c, \\[6pt]
c, & \text{if } y^{*}_{i} \leq c.
\end{cases}
\]

Thus, $y_{i}$ is always observed, while the latent outcome $y_{i}^{*}$ is only partially observed. The latent variable is modeled as
\begin{equation}\label{eq:2.1}
y_{i}^{*} = \boldsymbol{\Tilde{x}}_{i}^{\top} \boldsymbol{\alpha}_{0} + w_{i}\gamma_{0} + \epsilon_{i},
\end{equation}
where $\boldsymbol{x}^{*}_{i} = (\boldsymbol{\Tilde{x}}_{i}, w_{i})^{\top}$ denotes the $(p+1)$-dimensional vector of explanatory variables.  
Here, $\boldsymbol{\Tilde{x}}_{i} = (\Tilde{x}_{i1},\ldots,\Tilde{x}_{ip})^{\top}$ is a $p$-dimensional vector of exogenous regressors, with the first component normalized to one (intercept term). These covariates are assumed to be uncorrelated with the error term $\epsilon_{i}$. The variable $w_{i}$ represents an endogenous regressor that is correlated with $\epsilon_{i}$. For simplicity, we restrict attention to a single endogenous variable, but the methodology and results readily extend to cases with multiple endogenous regressors.  
The unknown parameter vector associated with the exogenous regressors is denoted by $\boldsymbol{\alpha}_{0} = (\alpha_{10}, \ldots, \alpha_{p0})^{\top}$, while $\gamma_{0}$ represents the coefficient of the endogenous variable. The error term $\epsilon_{i}$ is assumed to be independent of $\boldsymbol{\Tilde{x}}_{i}$. 
The censored regression model with left censoring at the fixed threshold $c$ can be expressed as
\begin{equation}\label{eq:2.2}
y_{i} = \max \{c, \, \boldsymbol{\Tilde{x}}_{i}^{\top} \boldsymbol{\alpha}_{0} + w_{i}\gamma_{0} + \epsilon_{i}\}, 
\quad i = 1,\ldots,n,
\end{equation}
commonly referred to as the Tobit model. Since no distributional assumption is imposed on $\epsilon_{i}$, this specification is often called a semi-parametric censored regression model.  
Without loss of generality, the censoring threshold can be normalized to zero ($c=0$), in which case the model simplifies to
\begin{equation}\label{eq:2.3}
y_{i} = \max \{0, \, \boldsymbol{\Tilde{x}}_{i}^{\top} \boldsymbol{\alpha}_{0} + w_{i}\gamma_{0} + \epsilon_{i}\}, 
\quad i = 1,\ldots,n.
\end{equation}
\section{Estimation Technique} In this section, we will discuss several estimation techniques that were introduced to estimate the parameter of the model defined in \eqref{eq:2.3}. We start with \cite{Nelson} and \cite{Amemiya1979}, who applied two-stage least squares ideas to the censored regression framework. Next,  the control function technique by \cite{Smith} is discussed, which has become a more generally used and computationally feasible solution.

Following that, the semiparametric and distribution-free estimators such as \cite{HONORE2004293} and \cite{BLUNDELL200765} for semiparametric censored regression are described. Finally, the  recent advancements, \cite{CHERNOZHUKOV2015201}, \cite{CHEN201830} and \cite{kobayashi} discuss the quantile-based techniques  and robust estimating frameworks  that deal with heavy-tailed errors and model misspecification.

\subsection{Classical Approaches}
The journey of estimation in censored regression models commenced by   \cite{Nelson} and \cite{Amemiya1979}, who applied traditional two-stage approaches, such as two stage least squares (2SLS). Their work provided the groundwork by demonstrating that standard Tobit estimators become inconsistent under endogeneity and proposing instrumental-variable-based adjustments.

Soon after, \cite{Smith} improved on these methods and  introduced a control function approach that addresses endogeneity by including residuals from the first-stage regression into the Tobit model. This methodology became a cornerstone because it was more adaptable and computationally practical than previous methods. The two techniques are described  as follows.

\subsubsection{Nelson and Olsen’s (1978) and Amemiya’s (1979) Estimation Techniques}
\cite{Nelson} proposed an instrumental variables approach for censored regression models with endogenous regressors. Note that the  instrumental variables are defined as the set of variables which are highly correlated with the explanatory variables, in limit and uncorrelated with the random errors, in limit. The instrumental variables help in obtaining a consistent estimator of the parameters, see \cite{RAO2008}. Their estimator replaces the endogenous explanatory variable with its fitted value, obtained from the first-stage regression of $w_{i}$ on the instruments. Specifically, if $\hat{w}_{i}=\boldsymbol{z}_{i}^\top \hat{\boldsymbol{\delta}}_{n}$ is the predicted value from the least squares regression in \eqref{eq:2.4}, then the structural Tobit model is estimated as
\[
y_{i}=\max \big\{0, \, \boldsymbol{\Tilde{x}}_{i}^\top \boldsymbol{\alpha}_{0} + \hat{w}_{i}\gamma_{0} + \epsilon_{i}\big\}.
\]
In this procedure, the standard Tobit maximum likelihood estimator is applied after substituting $\hat{w}_{i}$ in place of the endogenous regressor.  

Building on this framework, \cite{Amemiya1979} introduced an alternative estimator designed to improve efficiency. This method first estimates the reduced-form Tobit model for the endogenous variable and then applies a generalized least squares (GLS) step to recover the structural parameters. By exploiting additional information from the reduced form, Amemiya’s estimator achieves greater asymptotic efficiency than the Nelson–Olsen approach.  Nevertheless, this estimator does not reach the efficiency of the full information maximum likelihood method. The key limitation is that, unlike models with fully observed dependent variables, maximizing the marginal likelihood of the reduced form in the censored setting does not generally produce asymptotically efficient estimates of the reduced-form parameters. Consequently, while Amemiya’s approach improves upon the two-step substitution method of Nelson and Olsen, it still remains less efficient than full maximum likelihood.
\subsubsection{Smith and Blundell’s (1986) Estimation Technique}
The estimation technique of \cite{Smith} is a two-step procedure that modifies the Tobit model to account for endogeneity. The first stage involves regressing the endogenous variable on the instruments, while the second stage incorporates the residuals from this regression into the structural Tobit model.  
Let $z_{1,i}$ denote the $i$-th observation on the instrumental variable corresponding to the endogenous regressor $w_{i}$ in model \eqref{eq:2.1}, for $i=1,\ldots,n$. In the first stage, $w_{i}$ is regressed on the instrument $z_{1,i}$ and the exogenous covariates $\boldsymbol{\Tilde{x}}_{i}$ as follows:
\begin{equation}\label{eq:2.4} 
w_{i} = z_{1,i}\delta_{1,0} + \boldsymbol{\Tilde{x}}_{i}^\top \boldsymbol{\delta}_{2,0} + \vartheta_{i}
= \boldsymbol{z}_{i}^\top \boldsymbol{\delta}_{0} + \vartheta_{i}, \quad i=1,\ldots,n,
\end{equation}
where $\boldsymbol{z}_{i}=(z_{1,i}, \boldsymbol{\Tilde{x}}^\top_{i})^\top$ is a $(p+1)$-dimensional vector of instruments, $\boldsymbol{\delta}_{0}=(\delta_{1,0}, \boldsymbol{\delta}_{2,0})^\top$ is the corresponding parameter vector, and $\vartheta_{i}$ is the first-stage error term. Assuming $\mathbb{E}[\vartheta_{i}|\boldsymbol{z}_{i}]=0$, the parameter vector $\boldsymbol{\delta}_{0}$ can be estimated by OLS as
\[
\hat{\boldsymbol{\delta}}_{n} = \arg\min_{\delta} \sum_{i=1}^n (w_i - \boldsymbol{z}_{i}^\top \delta)^2,
\]
with residuals
$
\hat{\vartheta}_{i} = w_{i} - \boldsymbol{z}_{i}^\top \hat{\boldsymbol{\delta}}$.
These residuals $\hat{\vartheta}_{i}$ capture the endogenous component of $w_{i}$ that is correlated with the structural error $\epsilon_{i}$. Suppose that
\[
\begin{pmatrix}
\vartheta_{i} \\
\epsilon_{i}
\end{pmatrix}
\sim \mathcal{N}_2 \Bigg(
\boldsymbol{0},
\begin{pmatrix}
\sigma_1^2 & \sigma_{12} \\
\sigma_{12} & \sigma_2^2
\end{pmatrix}
\Bigg),
\]
so that $\epsilon_{i}$ can be expressed as
\[
\epsilon_{i} = \rho_{10}\vartheta_{i} + \varepsilon_{i},
\]
with $\varepsilon_{i}$ independent of $\vartheta_{i}$. Substituting this decomposition into the structural equation yields the transformed Tobit model
\[
y_{i} = \max \{0, \, \boldsymbol{\Tilde{x}}_{i}^\top \boldsymbol{\alpha}_{0} + w_{i}\gamma_{0} + \hat{\vartheta}_{i}\rho_{10} + \eta_{i}\},
\]
where $\eta_{i}= \varepsilon_{i}+\rho_{10}(\vartheta_{i}-\hat{\vartheta}_{i})$ is a new error term.  Thus, in the second stage, the conditional likelihood is constructed by including the residuals $\hat{\vartheta}_{i}$ from the first stage as additional regressors. This yields consistent estimates of the parameters. However, as noted by \cite{Smith}, this two-step procedure is generally less efficient than the full maximum likelihood estimator, since the least squares estimates of the reduced-form parameters in the first stage do not exploit the restrictions implied by the structural model.
 
\subsection{Semiparametric and Nonparametric Approaches} \cite{HONORE2004293} contributed semiparametric estimation methods that avoided strong parametric distributional assumptions, making estimation more robust to misspecification.

Later, \cite{Blundell} advanced the field by proposing a semiparametric censored regression estimator that combined control functions with flexible modeling of the error structure. This was a major leap since it relaxed assumptions about normality and homoscedasticity, making it applicable to a wider range of empirical settings. These two techniques are described in detail below.

\subsubsection{Honore and Hu Estimation Technique} For the censored regression model under an independence restriction, \citet{HONORE1994241} constructed moment conditions using a pairwise comparison approach. Building on this, \cite{HONORE2004293} extended the idea to models with endogenous regressors, formulating analogous moment conditions. While these conditions are satisfied at the true parameter values, \cite{HONORE2004293} noted that they do not guarantee uniqueness; that is, the moment conditions may not be sufficient to identify the regression coefficients fully. Although originally motivated by panel data censored regression models with predetermined (but not strictly exogenous) explanatory variables, the same principles and insights extend naturally to cross-sectional models with endogenous regressors, providing a unified perspective across panel and cross-sectional contexts.
To formalize the idea, consider a panel censored regression model with individual-specific fixed effects $\Tilde{\alpha}_{i}$, given by
 $$y_{it}=\max\{0,\Tilde{\boldsymbol{x}}^\top_{it}+w_{it}\gamma_{0}+\Tilde{\alpha}_{i}+\epsilon_{it}\},\;i=1,\ldots,n\;\text{and}\;t=1,\ldots, T,$$
where $\epsilon_{it}$ and $\epsilon_{is}$ are i.i.d. conditional on $(\Tilde{\boldsymbol{x}}_{it},\Tilde{\boldsymbol{x}}_{is},\boldsymbol{z}_{it},\boldsymbol{z}_{is},\Tilde{\alpha}_{i})$ for two distinct time periods $t$ and $s$. Under the restriction that $w_{it}\gamma_{0}\geq 0$ with probability one, the following auxiliary variables can be defined
$$e_{its}(\boldsymbol{\beta}_{0})=\max\{-\Tilde{\boldsymbol{x}}_{it}^\top\boldsymbol{\alpha}_{0},-\Tilde{\boldsymbol{x}}_{is}^\top\boldsymbol{\alpha}_{0},\Tilde{\alpha}_{i}+\epsilon_{it} \},$$
and similarly,
$$e_{ist}(\boldsymbol{\beta}_{0})=\max\{-\Tilde{\boldsymbol{x}}_{it}^\top\boldsymbol{\alpha}_{0},-\Tilde{\boldsymbol{x}}_{is}^\top\boldsymbol{\alpha}_{0},\Tilde{\alpha}_{i}+\epsilon_{is} \}.$$
It follows that for any function $h_{1}$, the condition $\mathbb{E}[h_{1}(e_{its}(\boldsymbol{\beta}_{0})-h_{1}(e_{ist}(\boldsymbol{\beta}_{0}))|\Tilde{\boldsymbol{x}}_{it},\Tilde{\boldsymbol{x}}_{is},\boldsymbol{z}_{it},\boldsymbol{z}_{is}]=0$ holds and can be exploited to construct valid moment conditions. Furthermore, if $(\epsilon_{it},\epsilon_{is})$ is exchangeable conditional on $(\Tilde{\boldsymbol{x}}_{it},\Tilde{\boldsymbol{x}}_{is},\boldsymbol{z}_{it},\boldsymbol{z}_{is},\Tilde{\alpha}_{i})$, then additional moment conditions are available of the form 
$$\mathbb{E}[g(h_{1}(e_{its}(\boldsymbol{\beta}_{0})-h_{1}(e_{ist}(\boldsymbol{\beta}_{0})))|\Tilde{\boldsymbol{x}}_{it},\Tilde{\boldsymbol{x}}_{is},\boldsymbol{z}_{it},\boldsymbol{z}_{is}]=0$$
for any odd function $g.$ These moment conditions can be used to estimate $\boldsymbol{\beta}$ via the continuously updating Generalized Method of Moments (GMM) framework \citep{Hansen1996}, which provides a practical and flexible estimation strategy in the presence of censoring and endogeneity.

\subsubsection{Blundell and Powell (2007) Estimation Technique}
Their approach is based on a “distributional exclusion” restriction, which assumes that the conditional distribution of the structural error depends on the regressors and instruments only through a lower-dimensional control variable. This control variable is defined as the residual from the reduced-form regression—that is, the difference between the endogenous regressor and its conditional expectation given the instruments.
Formally, the first-stage reduced form can be expressed as
$w_{i}=h(\boldsymbol{z}_{i})+\vartheta_{i}\; i=1,\ldots,n,$
where $h(.)$ is an unknown (possibly nonparametric) function of the instruments $\boldsymbol{z}_{i},$ and $\vartheta_{i}$ 
denotes the first-stage error.

This assumption translates into a similar exclusion restriction for the conditional quantiles of the censored dependent variable, motivating a two-stage estimation procedure. In the first stage, the conditional expectation 
$h(.)$ is estimated nonparametrically, and the corresponding residuals are obtained. In the second stage, \cite{BLUNDELL200765} estimates the structural parameters using a weighted least squares regression of pairwise differences in estimated quantiles of the dependent variable on the corresponding differences in regressors. The estimator is given by
\begin{eqnarray*}
(\hat{\boldsymbol{\alpha}}_{n},\hat{\gamma}_{n})^\top&=&\left[\sum_{i<j}K_{\vartheta}\left(\frac{\hat{\vartheta}_{i}-\hat{\vartheta}_{j}}{h_{n}}\right)\hat{t}_{i}\hat{t}_{j}(\boldsymbol{x}_{i}-\boldsymbol{x}_{j})(\boldsymbol{x}_{i}-\boldsymbol{x}_{j})^\top\right]^{-1}\\
&& \times \sum_{i<j}K_{\vartheta}\left(\frac{\hat{\vartheta}_{i}-\hat{\vartheta}_{j}}{h_{n}}\right)\hat{t}_{i}\hat{t}_{j}(\boldsymbol{x}_{i}-\boldsymbol{x}_{j})(\hat{\boldsymbol{q}}_{i}-\hat{\boldsymbol{q}}_{j}),
\end{eqnarray*}
where $q_{i}=Q_{\uptau}(y_{i}|\boldsymbol{x}_{i},\boldsymbol{z}_{i})=q_{i}(\uptau),$ $\boldsymbol{x}_{i}=(\Tilde{\boldsymbol{x}}_{i}^\top,w_{i})$ and 
$Q_{\uptau}(\epsilon_{i}|\boldsymbol{x}_{i},\boldsymbol{z}_{i})=Q_{\uptau}(\epsilon_{i}|\vartheta_{i})=\lambda_{\uptau}(\vartheta_{i}).$
Here, $K_{\vartheta}(.)$ denotes a kernel function that integrates to one, vanishes outside a compact set, and satisfies the usual regularity conditions. The sequence $\{h_{n}\}$ is a bandwidth parameter tending to zero at an appropriate rate as $n \to \infty.$ The trimming term 
$\hat{t}_{i}$ is defined as 
$$\hat{t}_{j}=\omega(\hat{q}_{i}).\mathds{1}\{(\boldsymbol{x}_{i},\boldsymbol{z}_{i})\in \mathscr{B}\}$$
where, $\omega(\hat{q}_{i})=0$ unless $\hat{q}_{i}>a$ for some $a>0.$ 
This ensures that only pairs of observations in the uncensored region and within a compact support $\mathscr{B}$ are used in estimation, with sufficiently small differences in their control variables.
The estimation uses only those pairs of observations where both estimated quantiles lie in the uncensored region and the difference in the estimated control variables is sufficiently small.  
Although this approach provides a consistent semiparametric estimator, it relies on several strong assumptions: (i) the endogenous regressors must be continuously distributed, (ii) a valid distributional exclusion restriction must hold, and (iii) the nonparametric first stage must be estimated at an appropriate rate. As a result, the method cannot be directly applied in settings with censored or discrete endogenous variables. This limitation is particularly important in program evaluation studies, where treatment variables are often binary or discrete-valued.

\subsection{Quantile-Based Estimation Approaches}
Quantile-based approaches represented a substantial advancement in the literature. \cite{Chernozhukov} and \cite{CHEN201830}, in particular, extended the framework of Censored Quantile Instrumental Variable (CQIV) estimation, allowing researchers to capture heterogeneous impacts at different points of the outcome distribution rather than focusing simply on mean effects. This method provides a more complete knowledge of the influence of endogenous regressors under censoring, especially when effects vary across quantiles. The two techniques are described  as follows.

\subsubsection{Chernozhukov (2015) and Chen (2018) Estimation Technique}
 \cite{CHERNOZHUKOV2015201} extended the approach of Chernozhukov and Hong (2002) to handle endogeneity within a non-additive triangular system of equations. Despite this generalization, both methods rely on the existence of a control function, which imposes strong structural assumptions on the underlying model. As a result, if the true nature of endogeneity is misspecified, the control function approach may yield inconsistent estimates and misleading inference. Moreover, this approach requires the endogenous variables to be continuously distributed, thereby excluding cases with censored or discrete endogenous variables. This restriction is particularly limiting in program evaluation studies, where the endogenous treatment variable is often binary or multi-valued but discrete, rendering the control function approach generally inapplicable.

To address some of these limitations, \cite{CHEN201830} studied sequential estimation methods for censored quantile regressions. These methods have been shown to offer efficiency advantages: when the model is correctly specified, estimators based on the independence restriction are generally more efficient than those relying solely on quantile restrictions. Extensive numerical results illustrating this efficiency gain can be found in \cite{HONORE1994241} and \cite{HONORE2004293}.

\subsubsection{Chen and Wang (2020) Estimation Technique}
\cite{ChenandWang} introduced moment-based estimators for censored outcome models in the presence of endogenous regressor, relying on independence and quantile restrictions, respectively.
Under the full independence assumption on $\epsilon_{i}$, conditional on $(\Tilde{\boldsymbol{x}}_{i}, \boldsymbol{z}_i)$, we can construct valid moment conditions. For each quantile $\uptau \in (\uptau_l, \uptau_u)$, let $c_0(\uptau)$ denote the $\uptau$-th quantile of $\epsilon_i.$
The corresponding moment conditions are
\begin{equation}\label{eq:2.5}
\mathbb{E}[\mathds{1}\{\Tilde{\boldsymbol{x}}_{i}^\top \boldsymbol{\alpha} + c > 0\}\big(\mathds{1}\{y_i - \boldsymbol{x}_i^\top \boldsymbol{\theta }< c\} - \uptau\big)(\Tilde{\boldsymbol{x}}^\top, \boldsymbol{z}_{i}^\top)^\top] = 0, \quad \uptau \in [\uptau_l, \uptau_u],
\end{equation} Here $\boldsymbol{\theta}=(\boldsymbol{\alpha},\gamma)^\top$ and $\boldsymbol{x}_{i}=(\Tilde{\boldsymbol{x}}_{i},\boldsymbol{z}_{i})^\top.$ To handle heavy censoring, top-quantile moment conditions are imposed
\begin{equation}\label{eq:2.6}
\mathbb{E}[\big(\mathds{1}\{y_i - \boldsymbol{x}_i^\top \boldsymbol{\theta }< c\} - \uptau\big)(\Tilde{\boldsymbol{x}}^\top, \boldsymbol{z}_{i}^\top)^\top I_{\boldsymbol{x}^{o}}(\boldsymbol{x}_i)] = 0, \quad \uptau \ge \uptau_u^0,
\end{equation}
for some top quantiles $\uptau_u^0$ and $I_{\boldsymbol{x}^{o}}(\boldsymbol{x}_i)=\mathds{1}(\boldsymbol{x}_{i}\in \boldsymbol{x}^{o})$ restricts the sample to a bounded support to handle extreme censoring. The semiparametric estimation proceeds as follows:
\begin{enumerate}
\item  For a given $\gamma$, estimate $\boldsymbol{\alpha}$ via pairwise comparison:
\begin{equation}
\hat{\boldsymbol{\alpha}}(\gamma) 
= \arg \min_{\boldsymbol{\alpha}} \, \mathbb{S}_n(\boldsymbol{\alpha}, \gamma), \quad 
\mathbb{S}_n(\boldsymbol{\alpha}, \gamma) 
= \frac{1}{\binom{n}{2}} \sum_{i<j} 
s\Big(y_i(\gamma), y_j(\gamma), (\tilde{\boldsymbol{x}}_{i}-\tilde{\boldsymbol{x}}_{j})^\prime \boldsymbol{\alpha}\Big),
\end{equation}
where $y_i(\gamma) = \max\{y_i - w_{i}\gamma, \gamma^*\} - \gamma^*$, $\gamma^* = \max\{-\gamma, 0\}$, and $s(.)$ is the convex loss function of \cite{HONORE1994241}.
\item Estimate the quantiles of residuals $\epsilon_i$ using the Kaplan–Meier estimator:
\begin{equation}
\hat{c}(\hat{\boldsymbol{\alpha}}(\gamma), \gamma, \uptau) = \inf\{t : \hat{F}(\hat{\boldsymbol{\alpha}}(\gamma), \gamma, t) \ge \uptau\}.
\end{equation}
\item Define the objective function over all quantiles:
\begin{equation}
\mathbb{G}_n(\boldsymbol{\alpha}, \gamma, c) = \int_{\uptau_l}^{\tau_u^0} \|\mathbb{G}_{1n}(\boldsymbol{\alpha}, \gamma, c(\uptau), \uptau)\| d\uptau 
+ \int_{\uptau_u^0}^{\tau_{u}} \|\mathbb{G}_{2n}(\boldsymbol{\alpha}, \gamma, c(\uptau), \uptau)\| d\uptau,
\end{equation}
where $\mathbb{G}_{1n}$ and $\mathbb{G}_{2n}$ are the sample analogs of the moment conditions defined by 
\begin{equation}
\begin{split}
   \mathbb{G}_{1n}(\boldsymbol{\alpha}(\gamma), \gamma, c(\uptau))&=\frac{1}{n} \sum_{i=1}^{n}\mathds{1}\{\Tilde{\boldsymbol{x}}_{i}^\top \boldsymbol{\alpha} +\min\{\gamma,0\}+c > 0\}\big(\mathds{1}\{y_i - \boldsymbol{x}_i^\top \boldsymbol{\theta }< c\} - \uptau\big) \boldsymbol{z}_{i},\\
   \mathbb{G}_{2n}(\boldsymbol{\alpha}(\gamma), \gamma, c(\uptau))&=\frac{1}{n} \sum_{i=1}^{n}\big(\mathds{1}\{y_i - \boldsymbol{x}_i^\top \boldsymbol{\theta }< c\} - \uptau\big)\boldsymbol{z}_{i} I_{\boldsymbol{x}^{o}}(\boldsymbol{x}_i).
\end{split}   
\end{equation}
\item Estimate $\gamma_0$ and $\boldsymbol{\alpha}_0$ as:
\begin{equation}
\hat{\gamma} = \arg\min_\gamma \mathbb{G}_n(\hat{\boldsymbol{\alpha}}(\gamma), \gamma, \hat{c}(\uptau)), \quad 
\hat{\boldsymbol{\alpha}} = \hat{\boldsymbol{\alpha}}(\hat{\gamma}).
\end{equation}
\end{enumerate}
This procedure achieves the identification and consistent estimation of $(\boldsymbol{\alpha}_0, \gamma_0)$ through independence-based quantile moment conditions. Although point identification follows from general assumptions on the distribution of observables, these conditions are often difficult to validate in practice. Building on this, \cite{CHESHER20231700} introduces an alternative framework for censored outcome models with endogenous regressors under instrumental variable restrictions. The focus is on Tobit-type left censoring at zero, with extensions to stochastic censoring. In contrast to control function methods, this approach does not impose explicit structure on the process generating endogenous regressors. Consequently, the model may deliver either partial or point identification depending on instrument strength and the support of exogenous variables. The paper characterizes identified sets and develops inference procedures for scalar functions of partially identified parameters. An application to UK household tobacco expenditures demonstrates how the method allows inference on the effect of endogenous total expenditure under both minimal and stronger Gaussian assumptions and contrasts these results with those obtained from fully specified triangular models.
\subsection{Recent Developments and Extensions} 
Recent past has witnessed some novel approaches such as the Genya Estimation Technique which provide alternative frameworks to jointly tackle censoring and endogeneity in a more flexible and robust manner. Alongside these, significant progress was made in the development of robust estimation methods aimed at reducing the impact of outliers, accommodating heavy-tailed error distributions, and addressing model misspecification. Collectively, these advancements have enhanced the reliability and applicability of inference in censored regression models under practical data scenarios. two such estimation  techniques are presented below.  

\subsubsection{Genya Estimation Technique}
\cite{kobayashi} introduces the $p$-th Tobit quantile regression models with endogenous variables in a Bayesian framework. In this framework, the first-stage regression of the endogenous variable on the exogenous variables incorporates the assumption that the $\tau$-th quantile of the error term is zero. The residual from this regression is then included in the $p$-th quantile regression model, ensuring that the $p$-th conditional quantile of the new error term is zero. To account for the unknown distribution of the first-stage regression error, both parametric and semiparametric approaches are employed around the zero $\tau$-th quantile assumption. Since the value of $\tau$ is generally unknown, it is treated as an additional parameter and estimated from the data. The effectiveness of the proposed models is illustrated using simulated data and applied to real data on the labour supply of married women.

\subsubsection{Robust Estimation Technique} 
\cite{shukla2026mestimationcensoredregressionmodel} proposed a two-step estimation procedure for censored regression models with endogenous regressors using the control function approach. In the first step, the endogenous regressor is modeled via ordinary least squares under standard error assumptions. The residuals from this regression, denoted as $\hat{\vartheta}_{i},$ are then included in the second stage. The second stage employs an M-estimation technique to estimate the model parameters. Specifically, the M-estimator $\hat{\boldsymbol{\theta}}_{n}$ is defined as the solution to the estimating equation
$$
\sum_{i=1}^{n}\psi\Big(y_{i}-\max\{0,\hat{\boldsymbol{x}}_{i}^\top \boldsymbol{\theta}\}\Big)=0,
$$
where $\psi(.)$ is the derivative of a general loss function $\rho(.)$, $\hat{\boldsymbol{x}}_{i}=(\Tilde{\boldsymbol{x}}_{i},w_{i},\hat{\vartheta}_{i})^\top$, and $\boldsymbol{\theta}=(\boldsymbol{\alpha},\gamma,\rho_{1})^\top.$  
Under suitable regularity conditions defined in \cite{shukla2026mestimationcensoredregressionmodel}, the M-estimator is  consistent and asymptotically normal, i.e.,
$$\hat{\boldsymbol{\theta}}_{n} \xrightarrow{p} \boldsymbol{\theta}_{0},$$ 
and  
$$\sqrt{n}\,(\hat{\boldsymbol{\theta}}_{n}-\boldsymbol{\theta}_{0}) \xrightarrow{d} \mathcal{N}\big(0,\, \Sigma_{M}\big),$$
where $\boldsymbol{\theta}_{0}$ is the true parameter vector and $\Sigma_{M}$ is the asymptotic covariance matrix in Theorem~3.2 from \cite{shukla2026mestimationcensoredregressionmodel}.

Building on this framework, \cite{shukla2025lestimationapproachtobitmodels} introduced a robust alternative based on L-estimation. The first step of the procedure remains the same, while the second step replaces M-estimation with an L-estimator defined as
$$
\pi_{n}=\int_{0}^{1}\hat{\boldsymbol{\theta}}_{n}(\tau)\, J(\tau)\, d\tau,
$$
where $\hat{\boldsymbol{\theta}}_{n}(\tau)$ is the censored quantile regression estimator at quantile $\tau$, and $J(\tau)$ is a smooth weight function defined on a compact subinterval of $(0,1).$   
The L-estimator also satisfies consistency and asymptotic normality under standard conditions, i.e.,
$$
 \hat{\boldsymbol{\pi}}_{n} \xrightarrow{p} \boldsymbol{\pi}_{0}, 
$$ and 
$$\sqrt{n}\,(\hat{\boldsymbol{\pi}}_{n}-\boldsymbol{\pi}_{0}) \xrightarrow{d} \mathcal{N}\big(0,\, \Sigma_{L}\big),
$$ where $\Sigma_{L}$ is the corresponding asymptotic covariance matrix defined in Theorem~3.4 from \cite{shukla2025lestimationapproachtobitmodels}.  
These results extend naturally to nonlinear censored regression models with endogenous regressors. If the nonlinear regression function $\Tilde{g}(\cdot)$ is Lipschitz continuous in both parameters and regressors, then both  M- and L-estimators retain their consistency property (see, in \cite{Shukla2025}). In addition to these theoretical guarantees, the performance of the proposed methods has been evaluated through Monte Carlo simulations and an empirical application using the ``Affair'' dataset, demonstrating both finite-sample efficiency and robustness in practical contexts. In summary, the M- and L-estimation procedures offer complementary strengths: both ensure asymptotic reliability, simulations validate their finite-sample behavior, and real data applications confirm their practical relevance.
\section {Appliactions} In this section, we provide the application of Mroz data set available in wooldridge package (available at \url{https://github.com/swati-1602/CRLM-Real-data.git}) to demonstrate the application of our proposed estimators. This dataset focuses on the labor force participation of U.S. women and includes data for 753 individuals. The response variable of interest is the total number of hours worked per week. Notably, 325 out of the 753 women reported not working any hours, rendering the dependent variable left-censored at zero. This implies that the censoring probability (CP) is of approximately $0.43$.

In this dataset, the explanatory variables include a woman's years of education (\textit{educ}), years of experience (\textit{exper}) and its squared term (\textit{expersq}), age (\textit{age}), the number of young children under six years of age (\textit{kidslt6}), and non-wife household income (\textit{nwifeinc}). Non-wife household income is treated as an endogenous variable due to its potential correlation with unobserved household preferences that may influence the wife's labour force participation.

To address this endogeneity, the husband's years of education (\textit{huseduc}) is employed as an instrumental variable. It is assumed that while this variable does not directly affect the wife's decision to participate in the labour force, it influences both the husband's income and the non-wife household income. The variable \textit{hours} represents the total number of hours a woman worked in a year. 

 In the context of this study, an exogenous variable is one that is uncorrelated with the error term. In contrast, an endogenous variable is correlated with the error term, typically due to omitted variables, simultaneity, or measurement error. This correlation can lead to biased and inconsistent parameter estimates if ignored. To demonstrate the impact of endogeneity, two analyses are conducted:

 \textbf{1) Exogenous Case :} A censored regression model (Tobit model) is estimated assuming all regressors are exogenous. The model parameters are obtained by minimizing the absolute loss function using the M-estimation framework, also known as the CLAD (Censored Least Absolute Deviations) method \cite{POWELL1984}.
 
  \textbf{2) Endogenous Case (Control Function Approach):}
  To account for endogeneity in the \textit{nwifeinc} variable, a control function approach is adopted.
  \begin{itemize}
      \item In the first stage, the potentially endogenous regressor (\textit{nwifeinc} is regressed on all exogenous variables and an instrumental variable (\textit{huseduc}).
      \item The residuals from this first-stage regression are obtained and included as an additional regressor in the censored regression model.
      \item The model is then re-estimated using the least absolute loss-based M-estimation method.
  \end{itemize}
\begin{table}[htbp!]
\centering
\caption{Parameter Estimates and Bootstrap Mean Squared Error (BMSE) for Exogenous and Endogenous Case.}
\begin{tabular}{|c|c|c|ccccc|}
\toprule
\textbf{Estimator} & \textbf{Variable} & \textbf{Estimate} & \multicolumn{5}{c|}{\textbf{BMSE}} \\
\cmidrule(lr){4-8}
 &  &  & $\boldsymbol{B=100}$ & $\boldsymbol{B=200}$ & $\boldsymbol{B=300}$ &$\boldsymbol{B=400}$ &$\boldsymbol{B=500}$\\
\midrule
& age & -0.0646& 0.5113 &  0.3960 &  0.3593 & 0.3325 & 0.2959 \\
& educ&-0.0100&  0.4840 & 0.3710 & 0.3348 & 0.3132 &  0.2736 \\
& exper&0.0253&   0.4694 & 0.3579 &  0.3220 & 0.3038 & 0.2623\\
\textbf{CLAD}& expersq   & 0.1437 & 0.4390 &0.3325 & 0.2977&0.2908 & 0.2428   \\
&  kidslt6  & -0.0530 &  0.5050 & 0.3902 & 0.3536  & 0.3279 & 0.2906\\
& kidsge6 & -0.0075 &  0.4828 & 0.3700 &  0.3338 &  0.3124 & 0.2727\\
& nwifeinc & -0.0107&   0.4843 & 0.3713 & 0.3350 & 0.3134 &  0.2738\\
\midrule
& age & -0.4872&0.2081 &0.1870& 0.1688&  0.1636 &0.1540  \\
& educ&0.3316& 0.1914 & 0.1857 & 0.1644& 0.1543&0.1261\\
& exper&-0.2086& 0.1269 &0.0837& 0.0801& 0.0741 & 0.0627\\
\textbf{IVCLAD}& expersq   & 2.0238& 0.0699& 0.0677& 0.0638&0.0633&0.0522\\
&  kidslt6 &-0.3742 & 0.0285& 0.0282& 0.0274& 0.0226& 0.0217  \\
& kidsge6 &-0.5716 & 0.2252&  0.2239& 0.2238& 0.2533&  0.2528  \\
& nwifeinc  &-0.3465&0.0937& 0.0800&  0.0795& 0.0794& 0.0754  \\
& res &  -0.3750&0.2071& 0.2005&0.1674&0.1609&0.1532\\
\bottomrule
\end{tabular}
\label{table:1}
\end{table}
The performance of both models is compared using the Bootstrap Mean Squared Error (BMSE) of the parameter estimates. The results show that the BMSE values under the control function approach are slightly smaller than those obtained under the exogeneity assumption, indicating an improvement in estimation efficiency and bias reduction when endogeneity is properly addressed.
This application thus highlights that ignoring endogeneity in censored models can lead to biased inference, whereas incorporating a control function correction within the M-estimation (CLAD-type) framework provides more reliable parameter estimates.

\section{Available Software for Estimation of Censored Regression Models} In practice, several statistical software packages provide functions and routines for estimating censored regression models, thereby facilitating empirical applications and replication of research findings. Among these, the censReg package \cite{henningsen2011censreg} is one of the most widely used in R. It estimates Tobit and censored regression models using maximum likelihood estimation (MLE) for cross-sectional data, and also supports random-effects MLE for panel data through Gauss–Hermite quadrature. The package accommodates both left- and right-censoring and provides tools for model diagnostics and inference.

Similarly, the AER package \cite{kleiber2008AERbook, kleiber2009AERpackage} offers the function tobit() for estimating Tobit models, while the cenmle() function from the NADA package serves as a front-end interface to the survreg() function in the survival package. The tobit() function in the VGAM package, on the other hand, performs estimation using its own maximum likelihood routine, rather than relying on survreg(). Additionally, the survival package provides the survreg() function, which can be adapted to fit censored regression models within the broader framework of survival analysis.

For models that involve endogeneity, the sampleSelection package provides both two-step and maximum likelihood estimators for handling sample selection and endogenous censoring models. When the focus extends to semiparametric or quantile-based estimation, the quantreg package becomes particularly useful. It includes the function crq(), which is suitable for estimating quantile regression models with either random or fixed censoring. The quantreg package offers flexible computational algorithms for quantile regression, and with appropriate modifications, these methods can be effectively adapted to censored data settings. Moreover,  the function MCMCtobit from the MCMCpack package uses the Bayesian Markov Chain Monte Carlo (MCMC) method to estimate
censored regression models.

In the present study, the quantile loss (L-estimation) approach has been implemented using user-defined optimization routines in R, while M-estimation has been carried out using the optim() function in R (). These implementations enable the estimation of parameters in censored Tobit models under endogeneity. The proposed approach complements the existing software ecosystem by providing a robust alternative to traditional likelihood-based methods, particularly in cases where the error distribution is non-normal or heavy-tailed. The R codes of the Real data study is available at \url{https://github.com/swati-1602/CRM.git}

At present, there is no R package that directly supports both M-estimation and L-estimation techniques for censored regression models. Hence, the development of an efficient algorithm and its incorporation into an R package for handling such estimation methods could form a valuable direction for future research.

\section{Conclusion}
This survey article provides an overview of development of various methodologies available in the literature to handle endogeneity in censored regression models. The main contribution is to compile   the  theoretical background and estimation techniques. The paper comprehends the advantages, disadvantages, and constraints of these estimation techniques. Such a development will facilitate further research in the area of censored regression with endogeneity and extend help to address its further extension to nonlinear, panel, and high-dimensional settings.

\noindent\bf{Acknowledgement} The authors would like to thank the editor-in-chief and a reviewer for sharing comments, which significantly improved the article.
\bibliographystyle{apalike}
\bibliography{Reference} 
\end{document}